\documentclass[runningheads]{llncs}
\usepackage{graphicx}
\usepackage{wrapfig,lipsum,booktabs}

\usepackage[acronym]{glossaries}
\usepackage{dsfont}
\usepackage{mathtools}
\usepackage{amsmath}
\usepackage{amssymb}
\usepackage{xcolor}
\usepackage{subfig}

\newcommand{\todo}[1]{}
\renewcommand{\todo}[1]{{\color{red} TODO: {#1}}}
\newcommand{\comment}[1]{}

\newcommand{\todelete}[1]{%
}

\newcommand{\tokeep}[1]{%
#1
}

\begin{document}

\newacronym{btcvae}{$\beta$-TCVAE}{$\beta$-Total Correlation Variational AautoEncoder}
\newacronym{bvae}{$\beta$-VAE}{$\beta$-Variational AautoEncoder}
\newacronym{vae}{VAE}{Variational AautoEncoder}

\title{Multi-scale Microaneurysms Segmentation Using Embedding Triplet Loss}
%
%
\author{Mhd Hasan Sarhan\inst{1,2}\orcidID{0000-0003-0473-5461} \and
Shadi Albarqouni\inst{1}\orcidID{0000-0003-2157-2211} \and
Mehmet Yigitsoy\inst{2}\orcidID{0000-0001-6598-0933}  \and
Nassir Navab\inst{1,3} \and
Abouzar Eslami\inst{2}\orcidID{0000-0001-8511-5541}}


%
\authorrunning{M.H Sarhan et al.}
%
\institute{Computer Aided Medical Procedures, Technical University of Munich, Germany \and
Carl Zeiss Meditec AG, Munich, Germany \and
Computer Aided Medical Procedures, Johns Hopkins University, Baltimore, USA}

\maketitle              

\begin{abstract}
Deep learning techniques are recently being used in fundus image analysis and diabetic retinopathy detection. Microaneurysms are an important indicator of diabetic retinopathy progression. We introduce a two-stage deep learning approach for microaneurysms segmentation using multiple scales of the input with selective sampling and embedding triplet loss.
The model first segments on two scales and then the segmentations are refined with a classification model. To enhance the discriminative power of the classification model, we incorporate triplet embedding loss with a selective sampling routine.
The model is evaluated quantitatively to assess the segmentation performance and qualitatively to analyze the model predictions. 
This approach introduces a $30.29\%$ relative improvement over the fully convolutional neural network.
\keywords{Deep Learning  \and Segmentation \and Ophthalmology}
\end{abstract}


%

\section{Introduction}
\label{sec:intro}
Diabetic retinopathy (DR) is the leading cause of vision impairment and blindness for middle-aged groups \cite{lee2015epidemiology}. DR early detection is important for the treatment planning. \tokeep{Severity of DR falls into one of five levels (none, mild, moderate, severe, or proliferative) \cite{drscale}}. Microaneurysms are considered as the first signs for detecting early stages of DR.
Hence, detecting these lesions is important for Computer Aided Diagnosis systems. Microaneurysms are abnormalities in the microvascular structure and appear as small red dots in color fundus images. Screening programs use colored fundus images of the retina for their rich information and ease of access. Detecting microaneurysms in colored fundus images is a challenging task due to the small size of the lesion which makes up less than 1\% of the entire image \cite{gargeya2017automatedclassification}, and the low contrast between microaneurysms and background.

Microaneurysms are the strongest determinant for DR since they are the first lesion that appears during the early stages. 
Various approaches for microaneurysms detection using deep learning are proposed \cite{haloi2015improved,orlando2017ensemble,lam2018patches}. These methods are patch-wise approaches and use deep architectures to extract representative features. These features could be added to a set of hand-crafted features \cite{orlando2017ensemble} and passed to a classification model or used solely in an end-to-end network \cite{haloi2015improved,lam2018patches}. Deep learning techniques in the literature of microaneurysms detection use random patches selection, hence, they are prone to be biased towards the oversampled class. Moreover, no work in the microaneurysms segmentation context has leveraged the embedding space of the input patches to impose an additional constraint on the learning process. 
\paragraph{Contributions: }In this work, a multi-scale patch-wise approach for segmenting microaneurysms in retinal fundus images is proposed. The main contributions of this work are 1) fusing segmentation on multiple scales for microaneurysms detection, and 2) using embedding triplet loss \cite{schroff2015facenet} with selective sampling \cite{van2016selective} to increase the descriptiveness of the feature representation while focusing the training on informative examples. The model is agnostic to other lesions (i.e. the model differentiates between healthy and microaneurysm patches regardless of information about other lesions). Being agnostic to other lesions is important in such cases as it may be difficult to obtain an annotated dataset with all DR lesions annotated.

\section{Methodology}
\label{sec:methodology}
\begin{figure*}[t]
\centering

    \includegraphics[width=0.9\textwidth]{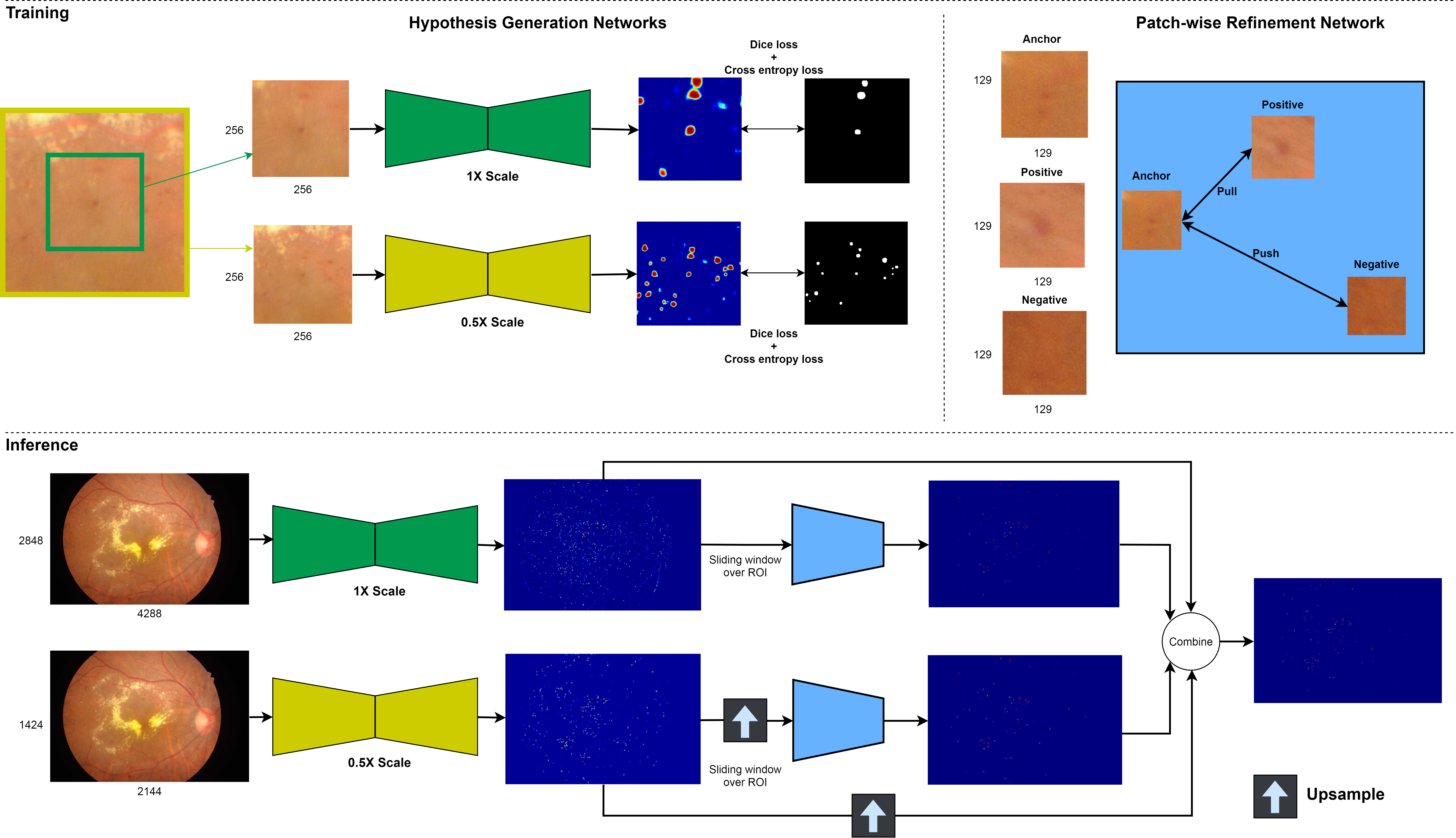} 

    \caption{Pipeline for the multi-scale microaneurysms segmentation framework. Top part shows the pipeline in the training mode where each model is trained separately. Bottom part shows the pipeline at inference time where the image is used rather than patches for the hypothesis generation networks.}
    \label{fig:pipeline}
\end{figure*}

Our proposed microaneurysms segmentation framework, depicted in Fig.~\ref{fig:pipeline}, consists of two stages; the \textit{hypothesis generation network} \textbf{(HGN)}, where multi-scale fully convolutional networks (FCNs) are employed to propose a region of interest (ROI), and \textit{patch-wise refinement network} \textbf{(PRN)}, where extracted patches around ROIs are passed to the classifier.
In the next sections we introduce the details of the applied method. First, we go through the fully convolutional hypothesis generation networks, the reasoning behind having multiple scales, and the details of the loss function used for optimization. The second section is dedicated for the PRN. In which, the motive behind this network is explained and the details of triplet loss and selective sampling are presented. 

\paragraph{Hypothesis Generation Network (HGN): }
High-resolution fundus images where a microaneurysm covers a very small part of the image are examined to segment microaneurysm. Using a zoomed-in patch would allow for high spatial accuracy on account of losing semantic information, whilst a zoomed-out patch would have a richer semantic representation on the account of losing spatial resolution \cite{ghiasi2016laplacian}. As a trade-off, we use equally sized patches on two scales of the image to build two HGNs, one for each scale.

HGN is a fully convolutional neural network trained on patches of size $256\times256$ extracted from the fundus images. Two HGNs are trained for two different scales of the fundus image (1x, 0.5x). This allows the extraction of scale-related features while at the same time preserve full resolution image information.
\tokeep{The architecture used is the full resolution residual network type A \cite{pohlen2017FRRN} for its good results in segmentation.}

To select the training patches, we define images that contain no signs of DR as healthy (negative) images and images with microaneurysms as lesion (positive) images. Healthy pixels are extracted only from healthy patients' scans and lesion pixels are extracted from DR patients at the microaneurysms locations. 
As a loss function, weighted cross entropy loss is used to compensate for the imbalance negative and positive patches. Moreover, dice loss is optimized to enhance the spatial overlap between a segmentation map output and the gold standard segmentation. We use a differentiable approximation of the dice loss as in~\cite{milletari2016vnet}.

\paragraph{Patch-wise Refinement Network}
PRN is a classification network that is used as on top of the HGN. The input of the network is an image patch and the output is the probability of the patch center pixel being a microaneurysm or healthy.
\tokeep{The segmentation maps} of the HGN are used as regions of interest for the PRN. The architecture of classification networks allows for receptive fields larger than fully convolutional networks that consume more memory because of the decoder part and skip connections. The larger receptive field allows for feature maps that incorporate more spatial information about the image which enriches the extracted features. The architecture employed for this network is an \tokeep{adopted} version of the Resnet-50 \cite{he2016resnet}. One downsampling step is omitted from the original architecture because the input image size in our case is smaller than what is expected in the Resnet-50 scenario.
In the training phase, patches are extracted from images in the same manner of extracting 1x resolution patches for HGN. The only difference is the size of PRN patches is $129\times 129$

To extract discriminative features in PRN we propose the utilization of triplet loss \cite{schroff2015facenet}. Triplet loss is applied on the embedding of a patch around pixel $x$ into a $d$-dimensional feature space. The aim of triplet loss is to make similar patches closer to each other in the embedding space while pushing dissimilar patches away from each other in the embedding space using a predefined distance measure. We \tokeep{found} the feature representation of the last convolution layer after the global average pooling (GAP) \tokeep{as a good representation in the embedding space due to its high descriptive power while having a compact representation.}
The optimization of triplet loss requires three input patches namely the anchor patch $x^a$, the positive patch $x^p$ and the negative patch $x^n$. The goal is to make the embedding of the positive patch closer to the anchor patch than the embedding of the negative patch. Patches with microaneurysms at the center pixels are used as anchor and positive patches, while healthy patches are used as negative patches. The loss is defined as

\begin{equation}
    \mathcal{L}_{triplet}= \sum^{N}_{i}\big[d(f(x_{i}^{a}), f(x_{i}^{p})) - d(f(x_{i}^{a}), f(x_{i}^{n})) + a\big]_{+}
    \label{eq:triplet}
\end{equation}
where $a$ is a margin to enforce a distance between positive and negative pairs, $d(.,.) \in \mathbb{R}^1$ is the distance measure in the embedding space, and $N$ is the number of all possible triplets. As a distance measure, angular cosine distance is utilized as 
it shows better performance on high dimensional representations when training deep networks \cite{nair2010rectified}.
\tokeep{In addition to triplet loss, cross entropy loss for patches is optimized.}

Generating all triplets, in this case, would be computationally prohibitive. Moreover, the imbalance in the dataset is high. To counter these problems, we use selective sampling~\cite{van2016selective}. This approach of training proved to enhance the results in training scenarios where data from different classes are not balanced. In our use case, the healthy class is over-represented. 
In selective sampling, patches with higher loss have a higher probability of being picked for the next epoch as they are considered representative samples.

\section{Experiments}
\subsection{Experimental Setup}
\label{sec:exp_setup}
\paragraph{Dataset}
For our evaluations \tokeep{of the segmentation pipeline}, we use the IDRiD\footnote{https://idrid.grand-challenge.org/} publicly available dataset. All images are captured with the same device that has 50-degree field of view and have size of $4288\times2848$ pixels. 
Before patch extraction, the published train dataset is split into two parts: training, and validation sets. The validation set is used for monitoring the training. Table \ref{tbl:dataset} shows the dataset splits.

\begin{table}[t]
    \centering
    \caption{IDRiD dataset splits}
    \label{tbl:dataset}
    \begin{tabular}{c|c|c||c|c}
         &  \multicolumn{2}{c}{\textbf{Healthy}} & \multicolumn{2}{c}{\textbf{Microaneurysms}}\\
         \hline
         & Images & Patches & Images & Patches\\
         \hline
         Train set& 80 & 6M & 44 & \textasciitilde500K\\
         Validation set & 9 & \textasciitilde6M & 10 & \textasciitilde132K\\
         Test set & 27 & - & 45 & -
    \end{tabular}
\end{table}

\paragraph{Implementation details}
We employ contrast enhancement following the formula 
$I_{pre}(x,y)=4 I(x,y) -4 G_{\sigma}*I(x,y) + 1024/30$.
To train HGN, we define a mini-batch of size 10 and consider each epoch to be 1000 mini-patches. 
The learning rate for the full-scale network is $1e-6$ and for the half-scale network is $1e-5$.

PRN is trained with mini-batches of triplets. The size of a mini-batch is $90\times3$ patches. We sample $90\times2$ patches from the pool of lesion patches randomly with uniform distribution, and $90\times1$ samples from the healthy patches pool with selective sampling. This neural network has a Siamese structure \cite{bromley1994siamese}, this means that each part of the triplet's three parts is run through identical versions of the network and the gradients are combined at the output to update the weights of the network.
In addition to triplet loss, cross entropy loss for pairs is optimized. To this end, we optimize the cross-entropy loss between the anchor and the negative pair. Every $1000$ mini-batches is considered as an epoch. 
We run selective sampling routine every $10$ epoch, this is because of the big number of training patches. which takes a significant amount of time to evaluate. Learning rate is set to $1e-5$ and decreased by a factor of $10$ after $20$ epochs. The optimization of the losses is done using Adam optimizer \cite{kingma2014adam}.

\subsection{Multi-scale effect}
\label{sec:multi-scale}
In this evaluation, we study the effect of using multiple scalse. To this end, two HGNs are trained, one for the full resolution image and one for the downsampled image by a factor of two. 
The evaluation is done on the publicly published test set images. We compare results from each scale with the results of combining the two scales \todelete{with two combination options} \tokeep{in two different ways}. First the output of the half scale is upsampled using linear interpolation, then the \tokeep{prediction maps} of the two scales are combined either with \tokeep{pixelwise} arithmetic or geometric averaging. The results of this evaluation are presented in Table. \ref{tbl:prn}. FCN 1x, FCN 0.5x represent the evaluation on the prediction map of the full scale and half scale HGNs, respectively. FCN geometric and FCN arithmetic refer to the results of combining the two scales with geometric and arithmetic averaging, respectively. The results show that combining the two scales gives better performance either way. We notice a higher recall from the half scale network but lowest precision, this reflects that the model is very sensitive to microaneurysms and generates a high number of false positives that drops down the overall performance. 

\subsection{Patch-wise refinement and triplet loss effect}
\label{sec:prn}
We evaluate the effect of 1) using a classification network to refine the classifications of HGNs and 2) using triplet loss in the classification network to refine HGNs results (i.e PRN).
To evaluate the classification network, we utilize patches from the image in a sliding window fashion and use the classification probability of each point to obtain segmentation maps. It is worth noting that sliding window does not go over all the image, but only the parts higher than a preset probability threshold ($0.5$ in our case) from HGN. Two segmentation maps will be obtained by sliding over the image masked with two HGNs outputs. Two prediction maps from two HGNs and two prediction maps from refining HGNs results with the classification networks combined as shown in Figure \ref{fig:pipeline}.

We first demonstrate the effect of incorporating a classification network to refine the results of HGNs. To this end, we use an edited version of PRN that uses only cross entropy loss without the triplet embedding  optimization. This network is denoted as \textit{cls}. Using the classification network on top of the fully convolutional networks enhances the results of the overall segmentation. The larger receptive field allows for more descriptive representations which in turn could suppress false positives that are triggered by HGNs. The effect of utilizing triplet embedding loss is then evaluated by training a PRN using triplets from the training set. We set the margin value $a$ from Equation \ref{eq:triplet} to $0.5$. At test time, this network is utilized in a sliding window fashion similar to \textit{cls}.  

Using triplet loss in a multi-scale approach with geometric averaging has an overall $30.29\%$ PR AUC improvement over the baseline fully convolutional neural network trained with weighted cross entropy. The improvement when incorporating triplet loss could be attributed to the quality of the learned representations where lesion patches are forced to be close to each other with a certain margin of difference from healthy ones.

Our results come in 4th place in the IDRiD challenge outdated leaderboard based on the metric used on the released test set. The challenge submission is currently closed. iFLYTEK-MIG used Mask-RCNN to segment 3 lesions at the same time. VRT used a U-net to segment four lesions all together. PATech used a patch-wise approach with false positives bootstrapping on lesions simultaneously. We notice that in all these models, information about lesions other than microaneurysms is utilized. This makes the disambiguation between lesion types (\textbf{e.g.} hemorrhages and microaneurysms) learned inherently in the model but has the drawback of requiring full annotation of multiple lesion types. The proposed model does not require information from other lesions to be trained.

\begin{table}[t]
    \centering
    \caption{Ablation test for emphasizing each part of the pipeline}
    \label{tbl:prn}
    
    \begin{tabular}{c|c|c|c|c}
    \centering
            & AUC PR & F1-score & Precision & Recall\\
        \hline
         HGN 1x - baseline&  0.3374 & 0.3618 & 0.2970 & \textbf{0.4626}\\
         HGN 0.5x & 0.3411 & 0.4001 & 0.4380 & 0.3682\\
         HGN geometric & 0.3622& 0.3866 & 0.5115 & 0.3108\\
         HGN arithmetic & 0.3701 &  0.4156 & 0.4741 & 0.3701 \\
         \textit{cls} geometric & 0.3895& 0.4153 & \textbf{0.5402} & 0.3374\\
         \textit{cls} arithmetic & \textbf{0.3905} &  \textbf{0.4368} & 0.4973 & 0.3895 \\
         \hline
         PRN arithmetic &0.3978& \textbf{0.4323} & 0.54051 & 0.3602\\
         PRN geometric & \textbf{0.4196} & 0.38477 & \textbf{0.61128} & 0.2807\\
         
         \hline
         \hline
         \tokeep{IDRiD iFLYTEK-MIG} &\textbf{0.5017}& - & - & -\\
         \tokeep{IDRiD VRT} & 0.4951& - & - & -\\
         \tokeep{IDRiD PATech} & 0.4740& - & - & -\\
    \end{tabular}
\end{table}

\subsection{Visual evaluation}
\begin{figure}[t]
    \centering
    \includegraphics[width=0.80\textwidth]{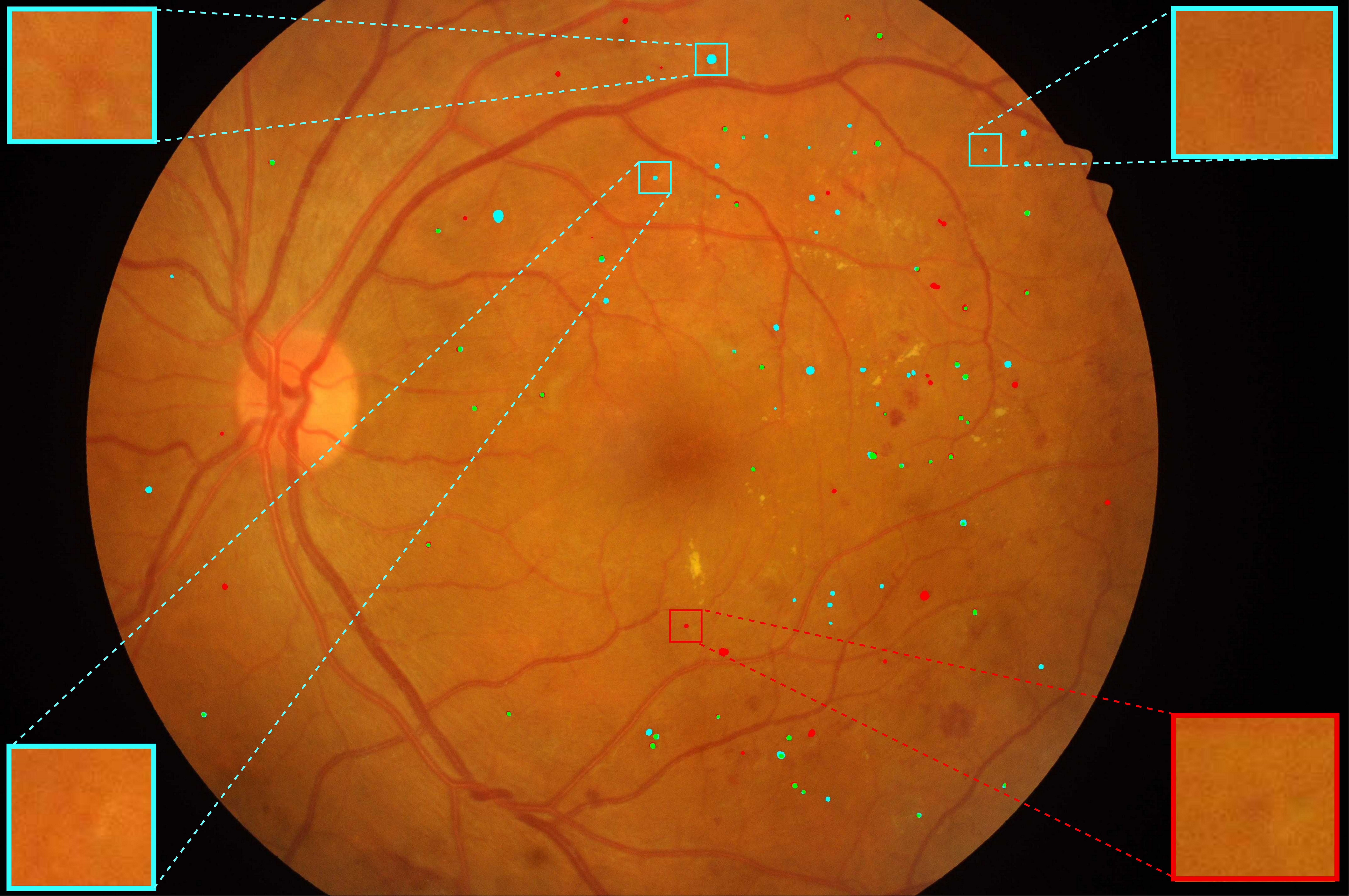} 
    \caption{\tokeep{An example of a segmented microaneurysms in a fundus image. Green is for true positives, red is for false positives, and cyan is for false negatives.}}
    \label{fig:example}
\end{figure}

We study the misclassifications of the model by visually examining samples of the results. In Figure. \ref{fig:example} an example of a segmentation is presented. From the example, we notice that false positives mostly lay in the area around hemorrhages or on top of a blood vessel where a higher intensity occur. False negatives are more difficult to be detected because they sometimes appear very close to hemorrhage and blend in or the contrast in the image is low enough to lose the microaneurysm. In the top left example, we see a false negative example where the microaneurysm is misclassified because of very light edges and irregular shape that leans towards hemorrhage. In the other false negative examples (in cyan), the cases are very difficult to be distinguished and variability between raters may occur in such cases. The bottom right example shows a false positive example where a darker area around the bright exudates appears similar to microaneurysm. The variability in illumination parameters of the capturing device has also a significant effect on the training and may lead to a bias towards a certain image appearance. It is important to note that images in the IDRiD dataset are compressed with a lossy compression which leads to big jumps in intensity values next to each other. For more examples please refer to supplementary material.

\section{Discussion and Conclusion}
\label{sec:discussion}
We hypothesize that using multiple fully convolutional networks for multiple scales of the inputs enhances the segmentation of small objects similar to microaneurysms because it gives a better trade-off between semantic and spatial accuracy. 
Embedding loss is employed mainly in learning image descriptors \cite{wohlhart2015learningdescriptor}. We use the triplet embedding loss in our model to treat deeper layers of the classification network as a local descriptor of the keypoint represented by the healthy or microaneurysm patch. The classification performance increases by adding this additional constraint on the features created by the network.

\tokeep{The segmentation results could be used in report generation for the doctors or in future studies to do big data analysis of populations. Microaneurysms turnover is also an important factor in the progression analysis of DR \cite{goatman2003turnover} and could be studied better with reliable models for microaneurysms segmentation.}


\bibliography{literature}

\end{document}